\documentclass[12pt]{article}

\usepackage{epsfig}
\usepackage{graphics}

\title{On     LAO Testing of   Multiple Hypotheses   for Many   Independent Objects\footnotetext{ The authors are with the Institute for Informatics and Automation Problems of the Armenian National Academy of Sciences, 1 P. Sevak str., Yerevan 0014, Armenia. E-mail: evhar @ipia.sci.am, par\_h@ipia.sci.am.}}

\author{\normalsize Evgueni A. Haroutunian {\it Associate Member, IEEE},  and  Parandzem M. Hakobyan}

\date{}

\newcommand{\df}{\stackrel{\bigtriangleup}{=}}

\begin{document}

\maketitle

\begin{abstract}
\vspace{2.3mm}

\setlength{\baselineskip}{12.7pt}
The problem of many hypotheses logarithmically asymptotically optimal (LAO) testing for a model consisting of three or more independent objects  is solved. It is supposed that $M$ probability distributions are known and each   object  independently of  others  follows to one of them. The matrix of  asymptotic interdependencies (reliability--reliability functions) of all possible pairs of the error probability exponents (reliabilities) in optimal testing for   this model is studied.

This problem was introduced (and solved for  the case of two objects and    two given probability distributions)  by   Ahlswede  and Haroutunian. The model with two independent objects with $M$  hypotheses  was explored  by Haroutunian and Hakobyan.
\end{abstract}

\begin{small}
\noindent {\bf Index Terms -} Hypothesis testing, multiple hypotheses,  logarithmically asymptotically optimal

\hspace{2.1cm}                   (LAO) tests,  two independent objects, error probability, reliability function.

\end{small}

\vspace{10mm}

\centerline{\rm \sc I. Introduction}
\vspace{5mm}
\noindent In \cite {2}
(see also \cite {R}, \cite {A}) Ahlswede and Haroutunian formulated an ensemble of new problems on multiple hypotheses testing for many objects and on identification of hypotheses. Noted problems are extentions of those investigated in the books  \cite {B1} and \cite {A1}.
Problems of  identification of  distribution and of   distributions ranking for one object  were solved in \cite {R} completely.  Also the problem of hypotheses testing for the model consisting  of two independent or two strictly dependent objects (when they cannot admit the same distribution) with  two possible hypothetical distributions was investigated in \cite {R}.
In this paper we study the model consisting of  $K(\geq 3)$  objects which independently follow to one of given   $M (\geq 2)$ probability distributions.  The  problem is a generalization of those  investigated in papers \cite {H1} -- \cite {B},  and  for  testing of many hypotheses  concerning one object  in  \cite {E5}. The case of two independent objects with three hypotheses was examined in  \cite {3}. Recently Tuncel \cite {T} published  an interesting consideration of the problem of multiple hypothesis optimal testing, which differs from the approach of   \cite {E5}, \cite {Hr 89}.

Let ${\cal P}({\cal X})$ be the space of all probability  distributions (PDs) on  finite set ${\cal X}$.  There are given $M$ PDs\,\,\, $G_m\in{\cal P}({\cal X})$, \,\,$m=\overline {1,M}$.

Let us recall main  definitions from \cite {E5} for the case of one object. The random variables (RV) $X$ taking  values on  $\cal X$  follows to one of the $M$ PDs $G_m$,  $m=\overline {1,M}$. The statistician  must accept   one of  $M$ hypotheses $H_m: G= G_m,\,\,\,\,m=\overline {1,M}$, on the base of  a sequence of results of $N$  observations of the object ${\bf x}=(x_1,...,x_n,...,x_N)$, $x_n\in {\cal X}$, $n=\overline {1,N}$. The procedure of decision making is a non-randomized test $\varphi_N$, which  can be defined by division of the sample space ${\cal X}^N$ on $M$ disjoint subsets ${\cal A}^N_l=\{{\bf x:}\,\,\, \varphi^N({\bf x})=l\}$, $l=\overline {1,M}$. The set ${\cal A}^N_l$ contains  all vectors $\bf x$ for which the hypothesis $H_l$ is adopted. The probability
$\alpha_{m|l}(\varphi_N)$ of the erroneous acceptance of hypothesis $H_l$ provided that $H_m$ is true, is equal to
$G^N_m(A_l^{N})$, $l\not=m$. The probability to reject $H_m$, when it is true, is
$$
\alpha_{m|m}(\varphi_N)\df\sum\limits_{l\not= m}\alpha_{m|l}(\varphi_N).\eqno (1)
$$

The error probability exponents   of the sequence of tests $\varphi$,  which it  is convienient to  call "reliabilities", are defined as
$$
E_{m|l}(\varphi)\df\overline {\lim\limits_{N\to \infty}}-\frac{1}{N}\log\alpha_{m|l}(\varphi_N),\,\,\,m,l=\overline {1,M}.\eqno (2)
$$
 It follows from (1) that
$$
E_{m|m}(\varphi)=\min_{l\not= m}E_{m|l}(\varphi),\,\,\,m=\overline {1,M}.\eqno (3)
$$
The matrix ${\bf E (\varphi)}=\{E_{m|l}(\varphi)\}$ is the reliability matrix of the sequence $\varphi$ of tests. It was studied in \cite {E5}.

\noindent {\it  Definition:}   { We call the sequence of tests $\varphi^*$ logarithmically asymptotically optimal (LAO) if for given positive values of $M-1$  diagonal elements of the matrix ${\bf E}(\varphi^*)$     maximal values to all other elements of it are provided}.

The concept  of LAO test was introduced by L. Birge \cite {B} and also elaborated in \cite {E5},  \cite {3} and \cite {Hr 89}.
Now let us  consider the model with   three objects. Let $X_1$, $X_2$ and $X_3$ be independent  RV  taking values in the same  finite set  ${\cal X}$ with one of $M$  PDs, they     are characteristics of corresponding  independent objects. The random vector $(X_1,X_2,X_3)$ assumes values $(x^1,x^2,x^3)\in {\cal X}\times {\cal X}\times {\cal X}$.

 Let $({\bf x_1},{\bf x_2},{\bf x_3})=((x_{1}^{1},x_{1}^{2},x_{1}^{3}),...,(x_{n}^{1},x_{n}^{2},x_{n}^{3}),...,(x_{N}^{1},x_{N}^{2},x_{N}^{3}))$, $x_n^k\in {\cal X}$, $k=\overline {1,3}$, $n=\overline {1,N}$, be a sequence of results of $N$ independent observations of the vector  $(X_1,X_2,X_3)$. It is necessary to   define  unknown PDs of the objects  on the base of   observed data. The decision for each object must be  made from   the same set of hypotheses: $H_m :G=G_m$, $m=\overline {1,M}$ .
We call  this  procedure  the test for three  objects and denote it by $\Phi_N$. It can be considered as  three  sequences of  tests $ \varphi^{1}$, $\varphi^{2}$, $\varphi^{3}$ by one for each object.   We will  denote the  compound test sequence   by ${{\Phi}}$. When we have $K$ independent objects the test  ${{\Phi}}$ is composed of $K$  sequences of  tests $ \varphi^{1}$, $\varphi^{2}$,..., $\varphi^{K}$.

Let   $\alpha_{m_1,m_2,m_3|l_1,l_2,l_3}(\Phi_N)$ be the probability of the erroneous acceptance by the test $\Phi_N$   of the hypotheses triple  $(H_{l_{1}},H_{l_{2}},H_{l_{3}})$ provided that the triple    $(H_{m_{1}},H_{m_{2}},H_{m_{3}})$ is true, where $(m_1,m_2,m_3)\not=(l_1,l_2,l_3)$, $m_i,l_i= \overline {1,M}$, $i=\overline {1,3}$. The probability to reject a true triple of hypotheses $(H_{m_{1}},H_{m_{2}},H_{m_{3}})$ by analogy with (1) is   the following:
$$
\alpha _{m_1,m_2,m_3|m_1,m_2,m_3}(\Phi_N)=\sum\limits_{(l_1,l_2,l_3)\not=(m_1,m_2,m_3)}\alpha_{m_1,m_2,m_3|l_1,l_2,l_3}(\Phi_N).\eqno(4)
$$
We  study corresponding   limits $E_{m_1,m_2,m_3|l_1,l_2,l_3}(\Phi)$ of  error probability exponents  of the sequence of tests $\Phi$, called  reliabilities
$$
E_{m_1,m_2,m_3|l_1,l_2,l_3}(\Phi)\df\overline{\lim\limits_{N\to\infty}}-\frac{1}{N}\log\alpha_{m_1,m_2,m_3|l_1,l_2,l_3}(\Phi_N),\,\,\,\,\,\,m_i,l_i= \overline {1,M},\,\,\,\,\,i={1,3}.\eqno {(5)}
$$
It follows from (5) that (compare with (3))
$$
E_{m_1,m_2,m_3|m_1,m_2,m_3}(\Phi)=\min\limits_{(l_1,l_2,l_3)\not=(m_1,m_2,m_3)}E_{m_1,m_2,m_3|l_1,l_2,l_3}(\Phi).\eqno (6)
$$

For the case of $K$ objects the error probability and reliability are considered also in papers \cite {2} -- \cite{A}. The   test sequence $\Phi^*$ is called LAO for the model with $K$ objects if for given positive values of certain  $K(M-1)$   elements  of the reliability matrix  ${\bf E}(\Phi^*)$ the procedure provides maximal values for all other elements of it.

Our aim in this paper is  to analyze     the reliability matrix ${\bf E}(\Phi^*)=\{E_{m_1,m_2,m_3|l_1,l_2,l_3}(\Phi^*)\}$ of LAO tests for three objects. The first idea was to study matrix ${\bf E}(\Phi)$ by renumbering the triples  of PD-s  from $1$ to  $M^3$. We can give $M^3-1$ diagonal elements of matrix ${\bf E}(\Phi)$ and apply Theorem 1. In this case the number of preliminary given elements of the matrix ${\bf E} (\Phi)$ will be grater  and the procedure of calculations will be longer than in the algorithm presented in Section 3.

The generalization of the problem  for $K$ independent objects will be discussed during the text  and in Section 4.

\vspace{10mm}

\centerline{\rm \sc II. LAO Testing of  Hypotheses for One Object}
\vspace{5mm}

 \noindent We define the  divergence (Kullback-Leibler distance) $D(Q||G)$  for  PDs  $Q,G \in {\cal P}({\cal X})$,  as usual (see \cite {81}):
$$
D( Q|| G)=\sum\limits_{x}Q(x)\log\frac{Q(x)}{G(x)}.
$$
We need to remind  the Theorem and its Corollaries from \cite{E5} for the convenience to have notations.

For given positive elements $E_{1|1}, E_{2|2},\dots, E_{M-1| M-1}$ we denote
$$
{\cal R}_l\df \{Q: D(Q||G_l)\leq E_{l|l}\}, \,\,\,\, l=\overline {1, M-1 },\,\eqno{(7.a)}
$$
$$
{\cal R}_{M}\df\{Q: D(Q||G_l)>E_{l|l}, \,\,\,\, l= \overline {1,M-1 }\}={\cal P}({\cal X})-\bigcup\limits_{l=1}^{M-1}{\cal R}_l,\eqno{(7.b)}
$$
and consider the following values:
$$
E_{l|l}^{\ast}=E_{l|l}^{\ast}(E_{l|l})\df E_{l|l}, \,\,\,\,l=\overline{ M-1 }, \eqno{(8.a)}
$$
$$
E_{m|l}^{\ast}=E_{m|l}^{\ast}(E_{l|l})\df \inf\limits_{Q\in{\cal
R}_l}D(Q||G_m), \,\,\,\, m= \overline {1, M },\,\,\, m\not=l,\,\,\, l= \overline {1, M-1},\eqno{(8.b)}
$$
$$ E_{m|M}^{\ast}=E_{m|M}^{\ast}(E_{1|1},\dots,E_{M-1|M-1})\df \inf\limits_{Q\in{\cal
R}_{M}}D(Q||G_{m}),\,\,\,\, m=\overline {1, M-1 }, \eqno{(8.c)}
$$
$$ E_{M|M}^{\ast}=E_{M|M}^{\ast}(E_{1|1},\dots,
E_{M-1|M-1})\df \min\limits_{l= \overline {1,M-1 }}E_{M|l}^{\ast}. \eqno{(8.d)}
$$

\noindent {\it  Theorem  1 \cite {E5}:} { If  the distributions $G_m$, $m= \overline {1, M}$, are different,  that is   all elements of the matrix $\{D(G_l||G_m)\}$,   are strictly positive,  then two statements hold:

a) when  the given numbers $E_{1|1}, E_{2|2},\dots, E_{M-1| M-1}$ satisfy conditions
$$
0< E_{1|1}<\min\limits_{l= \overline { 2,M}}D(G_l||G_1), \eqno (9.a)
$$

$$
 0< E_{m|m}<\min[\min\limits_{l= \overline {1, m-1 }}E_{m|l}^{\ast}(E_{l|l}), \min\limits_{l= \overline {m+1,M}}D(G_l||G_{m})],\,\,\,\, m=\overline { 2,M-1}, \eqno (9.b)
$$
then there exists a LAO sequence of tests $\varphi ^*$, the reliability matrix of which ${\bf E(\varphi ^*)}=\{E_{m|l}^{\ast}\}$ is defined in $(8)$ and all elements of it are strictly positive;

b) even if one of conditions $(9)$ is violated, then the reliability matrix of any such test includes  at least one element equal to zero (that is the corresponding error probability does not tend to zero exponentially)}.

\noindent {\it  Corollary 1 \cite {3}:} It can be proved that

$$
E_{m|m}^*=E_{m|M}^*,\,\,\,\,\,\,m=\overline {1,M-1}, \mbox {and}\,\,\,\,\,\,E_{m|m}^*\not= E_{m|l}^*, \,\,\,\,l\not=m,M.\eqno (10)
$$
\vspace{0.3cm}
\noindent {\it  Proof:}   Applying theorem of  Kuhn-Tucker in (8.b) we can derive that the elements $E_{l|l}^*$, $l=\overline {1, M-1}$ may be  determined by elements $E_{m|l}^*$, $m\not= l$, $m=\overline {1, M}$, by the  following inverse function
$$
E_{l|l}^*(E_{m|l}^*)=\inf\limits_{Q: D(Q||G_m)\leq E_{m|l}^*}D(Q||G_l).
$$
From  conditions (9) we see that $E_{m|m}^*$ can be equal  only to one among $E_{m|l}^*$,  $l=\overline {m+1, M}$.  Assume  that (10) is not true,  that is $E_{m|m}^*=E_{m|l}^*$, for $l=\overline {m+1,M-1}$. From  (8.b) it follows  that
$$
E_{l|l}^*(E_{m|l}^*)=\inf\limits_{Q: D(Q||G_m)\leq E_{m|l}^*}D(Q||G_l)=\inf\limits_{Q: D(Q||G_m)\leq E_{m|m}^*}D(Q||G_l)=E_{l|m}^*,
$$
$$
\,m=\overline {1, M-1},\,l=\overline {1, M-1}, m<l,
$$
but from  conditions (9) it  follows  that $E_{l|l}^*<E_{l|m}^*$ for $m=\overline {1, l-1}$. Our assumption is not correct, hence    (10) is valid.

\noindent {\it  Corollary 2 \cite {3}:} If one preliminary given   element   $E_{m|m}$, $m=\overline {1, M-1}$, of  the reliability matrix of an object is equal to zero, then  the corresponding elements of the matrix determined as functions of $E_{m|m}$, will be define as in the case of Stain's lemma \cite {81}:

$$
E'_{l|m}(E_{m|m})=D(G_m||G_l), \,\,\,\,\,\,l=\overline {1,M},\,\,\,l\not=m,\eqno(11)
$$
and the remaining elements of the matrix  are defined by  $E_{l|l}>0$, $l\not=m$, $l=\overline{1,M-1}$, as follows from Theorem 1.
\vspace{0.3cm}

\noindent {\it  Remark 1:} The number of elements $E_{m|m}$ equal to zero  may be any between 1 and  $M-1$. Generalization of Corollary 2 is straightforward.

\vspace{10mm}

\centerline{\rm \sc III. LAO Testing of Hypotheses  for Three Independent  Objects}
\vspace{5mm}
\noindent Now let us consider the case of three  independent objects and $M$ hypotheses.  The compound test $\Phi$ may be compoused from three separate tests  $\varphi^1$, $\varphi^2$, $\varphi^3$.

Let us denote by  $\bf E{(\varphi^i)}$ the reliability matrices  of the sequences of tests $\varphi^i$, $i=\overline {1,3}$, for each of the objects. The following  Lemma is a generalization of    Lemma from \cite {2} and \cite {3}.

\noindent {\it  Lemma 1:} { If  elements $E_{m|l}{(\varphi^i)}$, $m,l=\overline {1,M}$, $i=\overline {1,3},$ are  strictly positive,
then the following equalities hold for $\Phi=(\varphi^1,\varphi^2,\varphi^3)$:}
$$
E_{ m_1,m_2,m_3|l_1,l_2,l_3}(\Phi)=\sum\limits_{i=1}^3E_{m_i|l_i}(\varphi^i),\,\,\,\mbox{if}\,\,\,\, m_i\not=l_i,\,\,\,\, i=\overline {1,3},\eqno (12.a)
$$
$$
E_{ m_1,m_2,m_3|l_1,l_2,l_3}(\Phi)=\sum\limits_{i:}E_{m_i|l_i}{(\varphi^i)},\,\,\, m_{k}=l_{k},\,\, m_i\not=l_i,\,\,\,i\not=   k,\,\,i,k=\overline {1,3},\eqno (12.b)
$$
$$
E_{ m_1,m_2,m_3|l_1,l_2,l_3}(\Phi)=E_{m_i|l_i}{(\varphi^i)},\,\,\, m_{k}=l_{k},\,\, m_i\not=l_i,\,\,\,i\not= k, \,\,k,i=\overline {1,3}.\eqno (12.c)
$$
Equalities (12.a) are valid also if $E_{m|l}(\varphi^i)=0$ for several pairs  $(m,l)$ and several $i$.

\noindent{\it  Proof:}  It follows from the independence of the objects  that
$$ \alpha_{ m_1,m_2,m_3|l_1,l_2,l_3}(\Phi_N)=\prod\limits_{
i=1}^3\alpha_{m_i|l_i}(\varphi_{N}^i), \,\,\,\mbox{if}\,\,\,\, m_i\not=l_i, \eqno (13.a)
$$
$$
 \alpha_{ m_1,m_2,m_3|l_1,l_2,l_3}(\Phi_N)=(1-\alpha_{m_k|l_k} (\varphi_{N}^k)) \prod\limits_{
i\not=k} \alpha_{m_i|l_i}(\varphi_{N}^i),\,  \,\,\, m_{k}=l_{k},\,\, m_i\not=l_i,\,\,\,\,\,\, i,k=\overline {1,3}, \eqno (13.b)
$$
$$
 \alpha_{ m_1,m_2,m_3|l_1,l_2,l_3}(\Phi_N)= \alpha_{m_i|l_i}(\varphi_{N}^i)\prod\limits_{
i\not=k} (1-\alpha_{m_k|l_k}(\varphi_{N}^i)),\,\,\, m_{k}=l_{k},\,\, m_i\not=l_i,  k,i=\overline {1,3}.\eqno (13.c)
$$

Remark that here we    consider also  the  probabilities of right (not erroneous) decisions.
According to  the definitions (4) and (5)     from equalities (13) we obtain relations (12).

Now we shall show how we can  find LAO test from the set of  compound  tests $\{\Phi=(\varphi^1,\varphi^2,\varphi^3)\}$ when strictly positive elements   $E_{m,m,m|M,m,m}$, $E_{m,m,m|m,M,m}$ and $E_{m,m,m|m,m,M}$, $m=\overline {1,M-1}$, of the reliability matrix are given.

\noindent{\it   Lemma 2:} These  elements    can be strictly positive only in the following  three  subsets of tests $\{\Phi=(\varphi^1,\varphi^2,\varphi^3)\}$:

${\cal A}\df \{\Phi: E_{m|m}(\varphi^i)>0,m=\overline {1,M-1},\,\,\,i=\overline {1,3} \}$,

$
{\cal B}\df \{\Phi: \mbox { one or several}\,\,  m' \,\, \mbox { from } [1,M-1] \mbox { exist such that}\,\,E_{m'|m'}(\varphi^i)=0\,\,\mbox {for two\,\,i, }
$

$\,\,\,\,\,\,\,\,\,\,\,\,\,\,\,\,\,\,\mbox { but}\,\,E_{m'|m'}(\varphi^j)>0,\,\, i\not=j,\,\mbox { and for other}\,\,m<M,\,\,  E_{m|m}(\varphi^i)>0, \,i,j=\overline {1,3} \},
$

${\cal C}\df \{\Phi: \mbox { one or several}\,\,\,  m'\,\,\,\,  \mbox {from}\,\,\,  [1,M-1] \mbox { exist such that }\,\,\,\,\,E_{m'|m'}(\varphi^i)=0,\mbox { and}$

$\,\,\,\,\,\,\,\,\,\,\,\,\,\,\,\,\,\,\, \,\,\,\,\,\,\,\,\,\,\,\,\,\,\,\,\,\,\,\,\,\,\,\,\,\,\,\,\,\,\,\,\,\,\,\,\,\,\,\,\,\,\,\,\,\,\,\,\,\,\,\,\,\,\,\,\,\,\,\,\,\,\,\,\,\,\,\,\,\mbox { and for other}\,\,\,\, m<M,\,\,\,\,\,  E_{m|m}(\varphi^i)>0, \,\,\,i=\overline {1,3} \}.$

\vspace {1cm}
\noindent {\it  Proof :}
When $E_{m|m}(\varphi^i)>0$, then
$$
\overline{\lim\limits_{N\to \infty}}-\frac{1}{N}\log(1-\alpha_{m|m}(\varphi^i))=0,\,\,\,m=\overline {1,M},\,\,\,\,i=\overline {1,3}.
$$
and when $E_{m|m}(\varphi^i)=0$, then
$$
\overline{\lim\limits_{N\to \infty}}-\frac{1}{N}\log(1-\alpha_{m|m}(\varphi^i))>0,\,\,\,m=\overline {1,M},\,\,\,\,i=\overline {1,3}.
$$

From these equalities and inequalities, keeping in mind (5), (10)  and  (13) we obtain that the given elements are positive for $\Phi\in{\cal A}\bigcup{\cal B}\bigcup{\cal C} $. Proposition 1 is proved.
\vspace {3 mm}

Let  \,us  define  the  following family of decision  sets for given   positive elements  $E_{m,m,m|M,m,m}$, $E_{m,m,m|m,M,m}$ and $E_{m,m,m|m,m,M}$, $m=\overline {1,M-1}$:

$$
{\cal R}_m^{(i)}\df \{Q:D(Q||G_m)\leq E_{m,m,m|m_1,m_2,m_3}, \,\,m_i=M,\,\,m_j=m,\,\,\,i\not=j,\},\,\,\, m=\overline {1,M-1}\,\,\, i=\overline {1,3},
$$
$$
{\cal R}_{M}^{(i)}\df \{Q: D(Q||G_m)> E_{m,m,m|m_1,m_2,m_3},\,\,m_i=M, \,\,m_j=m,\,\,\,i\not=j,\,\,m=\overline {1,M-1}\}, \,\,\,i=\overline {1,3}.
$$

Let also
$$
E_{m,m,m|m,m,M}^*\df E_{m,m,m|m,m,M},
$$
$$
E_{m,m,m|m,M,m}^*\df E_{m,m,m|m,M,m},\eqno (14.a)
$$
$$
E_{m,m,m|M,m,m}^*\df E_{m,m,m|M,m,m,},
$$
$$
E_{ m_1,m_2,m_3|l_1,l_2,l_3}^*\df \inf\limits_{Q:Q\in R_{l_i}^{(i)}}D(Q||G_{m_i}),m_{k}=l_{k},\,\,m_i\not=l_i,i\not=k,\,\,i,k=\overline {1,3}, \eqno (14.b)
$$
$$
E_{ m_1,m_2,m_3|l_1,l_2,m_3}^*\df=\sum\limits_{i\not= k}\inf\limits_{Q:Q\in R_{l_i}^{(i)}}D(Q||G_{m_i}) , \,\,\,m_{k}=l_{k},\,\,m_i\not=l_i,\,\,i,k=\overline {1,3},\eqno {(14.c)}
$$
$$
E_{ m_1,m_2,m_3|l_1,l_2,l_3}^*\df E_{m_1,m_2,m_3|l_1,m_2,m_3}^*+E_{m_1,m_2,m_3|m_1,l_2,m_3}^*+E_{m_1,m_2,m_3|m_1,m_2,l_3},\,\,\,m_i\not=l_i. \eqno (14.d)
$$

The   result of the present paper is formulated in

\noindent {\it Theorem 2:} { If all  distributions  $G_m$, $m=\overline {1,M}$,  are different,   (and consequently $D(G_l||G_m)>0$,  $l\not=m$, $l,m=\overline {1,M}$), then the following  three statements are valid:

a)\,\,\,\,when   given strictly positive elements $E_{m,m,m|m,m,M}$, $E_{m,m,m|m,m,M}$ and $E_{m,m,m|M,m,m}$, $m=\overline {1,M-1}$,  meet  the following conditions
 $$
\max ( E_{1,1,1|M,1,1},E_{1,1,1|1,M,1},  E_{1,1,1|1, 1, M})<\min\limits_{l= \overline { 2,M }}D(G_l||G_1),\eqno {(15.a)}
$$
$$
E_{m,m,m |M,m,m}<\min[\min\limits_{l= \overline {1,m-1 }} E_{m,m,m|l,m,m}^*,\,\,\min\limits_{l= \overline{ m+1,M}}D(G_l||G_m)],\,\,m=\overline {2,M-1}, \eqno {(15.b)}
$$
$$
  E_{m,m,m|m,M,m}<\min[{\min\limits_{l= \overline {1,m-1}} E_{m,m,m|m,l,m}^*,\,\,\min\limits_{l= \overline { m+1,M }}D(G_l||G_m)}],\,\,m=\overline {2,M-1},\eqno {(15.c)}
$$
$$
  E_{m,m,m|m,m,M}<\min[{\min\limits_{l= \overline {1,m-1}} E_{m,m,m|m,m,l}^*,\,\,\min\limits_{l= \overline { m+1,M }}D(G_l||G_m)}],\,\,m=\overline {2,M-1},\eqno {(15d)}
$$
 then
 there\,\, exists \,\,a \,\,\,\,LAO \,\,  \,\,test  \,\,sequence\,\,\,\, $\Phi^*\in {\cal A}$, \,\,\,\,the \,\,reliability \,\,\,matrix \,\,\,of \,\,which ${{\bf E}(\Phi^*)}=\{E_{ m_1,m_2,m_3|l_1,l_2,l_3}(\Phi^*)\}$ is defined in $(14)$ and all elements of it are positive,

b) \,\,\,\,when even  one of the inequalities $(15)$ is violated, then there exists at least one element of  the matrix ${{\bf E}(\Phi^*)}$ equal to $0$,

c) \,\,\,\,for  given strictly positive numbers  $E_{m,m,m|M,m,m}$, $E_{m,m,m|m,M,m}$, and $E_{m,m,m|m,m,M}$, $m=\overline {1,M-1}$ the reliability matrix ${\bf E}(\Phi)$ of  the tests  $\Phi$ from  the defined in Lemma 2 families  ${\cal B}$ and ${\cal C}$ necessarily contains elements equal to zero}.
\vspace {3mm}

\noindent {\it   Proof:} a)  Conditions   (15)  imply  that   inequalities analogous to  (9) hold simultaneously for the case of three objects.
Really,
 using equalities (10)   we can rewrite inequalities (9) for  three  objects   as follows:
$$
\max (E_{1|M}(\varphi^{1}), E_{1|M}(\varphi^{2}), E_{1|M}(\varphi^{3}))<\min\limits_{l=\overline { 2,M}}D(G_l||G_1),\eqno (16.a)
$$

$$
  E_{m|M}(\varphi^{i})<\min[\min\limits_{l=\overline {1,m-1}  } E_{m|l}^{*}(\varphi^{i}),\,\,\,\min\limits_{l=\overline {m+1,M}  }D(G_l||G_m)],\,\,\,\,\,\,\,i=\overline {1,3},\,\,\,\,\,\,m=\overline {2, M-1},\eqno (16.b)
$$
We shall prove, for example, the inequalities   $(16.b)$, for $i=2$ which are the consequence of the  inequalities (15.c).
Let us consider the tests $\Phi\in {\cal A}$  such that $E_{m,m,m|m,M,m}(\Phi)=E_{m,m,m|m,M,m}$    and $E_{m,m,m|m,l,m}(\Phi)=E_{m,m,m|m,l,m}^*$,  $l= \overline {1,m-1}$, $m=\overline {{1, M-1}}$.   The corresponding error probabilities $\alpha_{m,m,m|m,M,m}({\Phi_{N}})$ and $\alpha_{m,m,m|m,l,m}(\Phi_{N})$  are given as products defined by (13.c).
Because $\Phi\in {\cal A}$, then
 $$\overline{\lim\limits_{N\to\infty}}-\frac{1}{N}\log(1-\alpha _{m|m}(\varphi_{N}^i))=0,\,\,i=\overline {1,3}. \eqno (17)
$$
According to (5), (13.c) and (17)  we obtain that
$$
E_{m,m,m|m,M,m}^*(\Phi)=E_{m|M}^*(\varphi^{2}),\,\,\,\,m=\overline {2,M-1},
\eqno (18.a)
$$
$$
E_{m,m,m|m,l,m}^*(\Phi)=E_{m|l}^{*}(\varphi^{2}),\,\,\,\,m=\overline {2,M-1}.
\eqno (18.b)
$$
So (16.c) is consequence of (15.c).

 As we noted in the beginning of the proof  it follows  from (10) and (16) that conditions (9) of  Theorem 1 take place  for  each of  three objects. According to Theorem 1 there exists LAO sequences of tests $\varphi^{*,1}$ , $\varphi^{*,2}$ and $\varphi^{*,3}$  three  objects  such that the elements of the matrices $\bf E(\varphi^{*,i})$,  $i=\overline {1,3}$, are determined according to  (8). We  consider the sequence of tests   $\Phi^*$, which is composed of three  sequences of  tests $ \varphi^{*,1}$, $\varphi^{*,2}$, $\varphi^{*,3}$   and we will show that $\Phi^*$ is LAO and other elements of the matrix ${\bf E}(\Phi^*)$ are determined according to  (14).

It follows from (16), (10) and (9) that   the requirements of Lemma 1 are fulfilled. Applying   Lemma 1 we can deduce that the reliability matrix ${\bf E}(\Phi^*)$ can be obtained from matrices   $\bf E(\varphi^{*,i})$   as in (12).

  When  conditions (15) take place, we obtain (14) according to (12), (8), (10) and (18).  The  equality in (14.e) is a particular case of (6). From (14) it  follows that all elements of ${\bf E}(\Phi^*)$ are positive.

 Now it is easy to verify that the compound test $\Phi^*$ for three  objects  is  LAO.

b) When  one of the inequalities (15) is violated, then from (14.b) we see, that  some  elements in the matrix ${\bf E}(\Phi^*)$ must be equal to zero.

c) When  $\Phi\in {\cal B}$, then from (10) and (12.a) we can see that the elements $E_{m',m',m'|M,M,M}=\sum\limits_{i=1}^3E_{m'|M}(\varphi)=0$.

Let  $\Phi=(\varphi^1,\varphi^2,\varphi^3)\in {\cal C}$. For example $E_{m'|m'}(\varphi^1)>0$, $E_{m'|m'}(\varphi^2)=E_{m'|m'}(\varphi^3)=0$,    then
$$
E_{m',m',m'|m',M,M}=\overline{\lim\limits_{N\to \infty}}-\frac{1}{N}\log(1-\alpha_{m'|m'}(\varphi^1))+ E_{m'|m'}(\varphi^2)+E_{m'|m'}(\varphi^3)=0.
$$
Theorem 2 is proved.
\vspace {3mm}

\noindent {\it  Remark 2:} For every test  $\Phi$ from noted in Lemma 2 subset ${\cal C}$ (for the case of  subset ${\cal B}$ resonemets are similar) for given $3(M-1)$ elements of matrix ${\bf E}(\Phi)$, using independence of three objects,  the definition (4) and equalities (10) we can determine all other elements of ${\bf E}(\Phi)$ in the following way:

$$
E_{m',m',m'|M,m',m'}(\Phi)=\overline{\lim\limits_{N\to \infty}}-\frac{1}{N}\log(1-\alpha_{m'|m'}(\varphi^3))+\overline {\lim\limits_{N\to \infty}}-\frac{1}{N}\log(1-\alpha_{m'|m'}(\varphi^2))>0,
$$
$$
E_{m',m',m'|m',M,m'}(\Phi)=\overline{\lim\limits_{N\to \infty}}-\frac{1}{N}\log(1-\alpha_{m'|m'}(\varphi^1))+\overline{\lim\limits_{N\to \infty}}-\frac{1}{N}\log(1-\alpha_{m'|m'}(\varphi^3))>0,
$$
$$
E_{m',m',m'|m',m',M}(\Phi)=\overline{\lim\limits_{N\to \infty}}-\frac{1}{N}\log(1-\alpha_{m'|m'}(\varphi^1))+\overline{\lim\limits_{N\to \infty}}-\frac{1}{N}\log(1-\alpha_{m'|m'}(\varphi^2))>0.
$$
From these equalities it follows that
$$
\overline{\lim\limits_{N\to \infty}}-\frac{1}{N}\log(1-\alpha_{m'|m'}(\varphi^1))=
$$
$$
=\frac{1}{2}[E_{m',m',m'|m',M,m'}(\Phi)+E_{m',m',m'|m',m', M}(\Phi)-E_{m',m',m'|M,m',m'}(\Phi)],\eqno (19.a)
$$
$$
\overline{\lim\limits_{N\to \infty}}-\frac{1}{N}\log(1-\alpha_{m'|m'}(\varphi^2))=
$$
$$
=\frac{1}{2}[E_{m',m',m'|M,m',m'}(\Phi)+E_{m',m',m'|m',m',M}(\Phi)-E_{m',m',m'|m',M,m'}(\Phi)],\eqno (19.b)
$$
$$
\overline{\lim\limits_{N\to \infty}}-\frac{1}{N}\log(1-\alpha_{m'|m'}(\varphi^3))=
$$
$$
=\frac{1}{2}[E_{m',m',m'|M,m',m'}(\Phi)+E_{m',m',m'|m',M,m'}(\Phi) -E_{m',m',m'|m',m',M}(\Phi)],\eqno (19.c)
$$
Hence
$$
E_{ m_1,m_2,m_3|l_1,l_2,l_3}(\Phi)= E_{m_1|l_1}(\varphi^1)+E_{m_2|l_2}(\varphi^2)+E_{m_3|l_3}(\varphi^3),\,\,\,\,\, m_i\not=l_i,\,\,\,\, \i=\overline {1,3},\eqno (20.a)
$$
$$
E_{ m_1,m_2,m_3|l_1,l_2,l_3}(\Phi)=\sum\limits_{i:i\not= k}E_{m_i|l_i}(\varphi^3), \,\,\,\mbox {if}\,\,\, m_i\not= l_i,\,\,\,m_k=l_k, k\not= i.\eqno (20.b)
$$
Denoting right sums from (19.a) by A,  from (19.b) by B, from (19.c) by C  we obtain that
$$
E_{ m',m_2,m_3|m',l_2,l_3}(\Phi)= E_{m_2|l_2}(\varphi^2)+E_{m_3|l_3}(\varphi^3)+  A,\,\,\,\, m_2\not=l_2,\,\,\,\,m_3\not=l_3,
$$
$$
E_{ m_1,m',m_3|l_1,m',l_3}(\Phi)= E_{m_1|l_1}(\varphi^1)+E_{m_3|l_3}(\varphi^3)+B,\,\,\,\,
m_1\not=l_1,\, m_3\not=l_3,\eqno (20.c)
$$
$$
E_{ m_1,m_2,m'|l_1,l_2, m' }(\Phi)= E_{m_1|l_1}(\varphi^1)+E_{m_2|l_2}(\varphi^2)+C,\,\,\,\,  m_1\not=l_1,\,\,\,m_2\not=l_2,
$$
$$
E_{ m',m',m_3|m',m',l_3}(\Phi)=E_{m_3|l_3}(\varphi^3)+E_{m',m',m'|m',m',M},
$$
$$
E_{ m',m_2,m'|m',l_2,m'}(\Phi)=E_{m_3|l_3}(\varphi^3)+E_{m',m',m'|m',M,m'},\eqno (20.d)
$$
$$
E_{ m_1,m',m'|l_1,m',m'}(\Phi)=E_{m_3|l_3}(\varphi^3)+E_{m',m',m'|M,m',m'}.
$$
Remember that  in this case the elements $E_{ m',m',m'|M,M,M}(\Phi)=0$.
From (20) we see, that  LAO test   $ \Phi'=(\varphi'^1,\varphi'^2,\varphi'^3) $, $m'\in [1,M-1]$ is composed of  the tests $\varphi'^1$, $\varphi'^2$ and $\varphi'^3$ discussed in   Corollary 2.

\vspace{10mm}

\centerline{\rm \sc IV.  Supplements for $K(>)3$ Objects}
\vspace{5mm}
When we consider  the model with $K$ independent objects the  generalization of   Lemma 1 will take  the  following form:

\noindent {\it Lemma 3:} { If  elements $E_{m|l}{(\varphi^i)}$, $m,l=\overline {1,M}$, $i=\overline {1,K},$ are  strictly positive,
then the following equalities hold for $\Phi=(\varphi^1,\varphi^2,..., \varphi^K)$:}

$$
E_{ m_1,m_2,...,m_K|l_1,l_2,...,l_K}(\Phi)=\sum\limits_{i=1}^KE_{m_i|l_i}(\varphi^i),\,\,\,\mbox{if}\,\,\,\, m_i\not=l_i,\,\,\,\, i=\overline {1,K},
$$
$$
E_{ m_1,m_2,...,m_K|l_1,l_2,...,l_K}(\Phi)=\sum\limits_{i:\,\,i\not=j}E_{m_i|l_i}(\varphi^i),\,\,\,\mbox{if}\,\,\,\, m_j=l_j,\,\,\,\,m_i\not=l_i,\,\,\,\, i\j,\,\,\,\,\, i,j=\overline {1,K}.
$$

 For  given $K(M-1)$   strictly positive elements $E_{m,m,...,m|M,m,...,m}$, $E_{m,m,...,m|m,M,...,m}$, ...., $E_{m,m...,m|m,...,m,M,}$, $m=\overline {1, M-1}$ for the case of $K$ independent objects. We can find  the LAO test $\Phi ^*$ we will find as in the case of three independent objects. So the problem of many hypotheses testing of the model with $K$ independent objects may be solved by constricting corresponding sets $R_m^{(k)},\,\,\,\,k=\overline {1,K}, \,\,\,\,\,\,m=\overline {1,M}$ as in (14) and formulating conditions analogical to (15).

 It is interesting to analyse the generalization of  Corollary 2 for the case of many objects.

\vspace{10mm}

\centerline{\rm \sc V. Example}
\vspace{5mm}

\noindent Let us consider an example on LAO testing hypotheses concerning one object and two objects.  The  set ${\cal X}=\{0,1\}$ of two elements  and the following probability distributions given on ${\cal X}$:
$G_1=\{0,10; 0,90\},$ $G_2=\{0,85; 0,14\},$ $G_3=\{0,23; 0,77\}$.
In Fig. 1 and Fig.  2   the results of calculations  of functions  $E_{2|1}(E_{1|1})$ and $E_{2,1|1,2}(E_{1,1|3,1},E_{2,2|2,3})$   are presented. For these  distributions we have  $\min(D(G_2,G_1), D(G_3,G_1))\approx 2,2$ and \\$\min(E_{2,2|2,1}, D(G_3,G_2))\approx 1,4$. We see that when  the first inequality in (9.a) is violated then $E_{2|1}=0$ and,
when the inequality (15.b) and (15.c)  are  violated, then $E_{2,1|1,2}=0$.
\newpage
\mbox{}

\vspace {-5cm}
\begin {center}
\hspace{ 2.8cm}
\begin{picture}(360, 320)
\special{wmf:Graphic6.wmf x=8cm, y=5cm}
\end{picture}
\end {center}
\vspace{-0.5cm} \centerline {Fig. 1}
\begin {center}
\vspace{2cm}
\begin{picture}(360, 320)
\special{wmf:Graphic5.wmf x=12.5cm, y=9cm, z=6cm}
\end{picture}
\end {center}
\centerline {Fig. 2}


\end{document}